\documentclass{aastex}
\usepackage{graphicx,epsfig,emulateapj5}

\newcommand{\etal}{{et al.~}}

\newcommand{\bq}{\begin{equation}}
\newcommand{\eq}{\end{equation}}

\def\gtsim{\lower.5ex\hbox{$\buSildrel > \over\sim$}}
\def\ltsim{\lower.5ex\hbox{$\buildrel < \over\sim$}}
\def\arcsec{^{\prime\prime}}
\def\arcmin{^\prime}
\def\farcm{\hbox{$.\!\!^{\prime}$}} 
\def\farcs{\hbox{$.\!\!^{\prime\prime}$}}

\def\sun{\mbox{$_\odot$}}
\def\apjl{ApJL}
\def\apj{ApJ}
\def\apjs{ApJS}
\def\mnras{MNRAS}
\def\araa{ARAA}
\def\aj{AJ}
\def\aap{A\&A}
\def\aaps{A\&A Suppl.}
\def\nat{Nature}

\slugcomment{Accepted by ApJ Letters. To appear in  Nov 2004 issue}
\shorttitle{GEMS:Bar Evolution over the Last 8 Gyr}
\shortauthors{Jogee {\it et al}}
\begin{document}
\title{Bar Evolution Over the Last  Eight Billion  Years: A Constant 
Fraction of Strong Bars in GEMS
}
\author{Shardha Jogee\altaffilmark{1}, 
Fabio D. Barazza\altaffilmark{1},
Hans-Walter Rix\altaffilmark{2},
Isaac Shlosman\altaffilmark{3},
Marco Barden\altaffilmark{2},
Christian Wolf\altaffilmark{4},
James Davies\altaffilmark{1},
Inge  Heyer\altaffilmark{1},
Steven V.W. Beckwith\altaffilmark{1,5},
Eric F. Bell\altaffilmark{2},
Andrea Borch\altaffilmark{2},
John A. R. Caldwell\altaffilmark{1},
Christopher J. Conselice\altaffilmark{6},
Tomas Dahlen\altaffilmark{1},
Boris H\"{a}ussler\altaffilmark{2},
Catherine  Heymans\altaffilmark{2},
Knud Jahnke\altaffilmark{7},
Johan H. Knapen\altaffilmark{8},
Seppo Laine\altaffilmark{9},
Gabriel M. Lubell\altaffilmark{10},
Bahram Mobasher\altaffilmark{1},
Daniel H. McIntosh\altaffilmark{11}
Klaus Meisenheimer\altaffilmark{2},
Chien Y. Peng\altaffilmark{12},
Swara Ravindranath\altaffilmark{1},
Sebastian F. Sanchez\altaffilmark{7},
Rachel S. Somerville\altaffilmark{1} and 
Lutz Wisotzki\altaffilmark{7}}

\altaffiltext{1}{Space Telescope Science Institute, 3700 San Martin Dr., Baltimore, MD 21218; jogee@stsci.edu}
\altaffiltext{2}{Max-Planck Institute for Astronomy, Koenigstuhl 17, 
 D-69117 Heidelberg, Germany}
\altaffiltext{3}{Department of Physics and Astronomy,University of Kentucky,  Lexington, KY 40506-0055}
\altaffiltext{4}{Astrophysics, University of Oxford, Keble Road, Oxford OX1 3RH, U.K.}
\altaffiltext{5}{Department of Physics and Astronomy, Johns Hopkins University,  Charles and 4th Street, Baltimore, MD 21218}
\altaffiltext{6}{Department  of Astronomy, California Institute of Technology, Pasadena, CA 91125}
\altaffiltext{7}{Astrophysikalisches Institut Potsdam, An der Sternwarte 16, D-14482 Potsdam, Germany}
\altaffiltext{8} {Department of Physical Sciences, University of Hertfordshire,  Hatfield, Herts AL10 9AB, U.K.}
\altaffiltext{9}{{\it Spitzer} Science Center, California Institute of Technology, Mail Code 220-6, 1200 East California Boulevard, Pasadena, CA 91125, U.S.A,}
\altaffiltext{10}{ Department of Physics and Astronomy, Vassar College, Poughkeepsie, NY 12604, U.S.A}
\altaffiltext{11}{Department of Astronomy, University of Massachusetts, Amherst, MA 01003, U.S.A}
\altaffiltext{12}{Department of Astronomy, University of Arizona, 933 North Cherry Avenue, Tucson, AZ 85721, U.S.A}

\begin{abstract}
One third  of  present-day spirals host optically visible strong 
bars that  drive their dynamical evolution.
However, the fundamental question of how bars evolve over cosmological 
times has yet to be resolved, and even the frequency of bars at 
intermediate redshifts remains controversial. 
We investigate the frequency of bars  out to  $z$~$\sim$~1  
drawing on a sample of 1590  galaxies from the Galaxy Evolution 
from Morphology and SEDs survey, which provides  
morphologies from \it Hubble Space Telescope \rm
Advanced Camera for Surveys  (ACS)  two-band images and accurate redshifts
from the COMBO-17 survey. We identify spiral galaxies using three 
independent techniques based on the Sersic index, concentration parameter, 
and rest-frame color. We characterize  bar and disk features  by fitting 
ellipses to F606W and F850LP images,  using the two bands to minimize 
shifts in the rest-frame bandpass. We exclude highly inclined  
($i >$~$60^\circ$) galaxies to ensure reliable morphological classifications 
and  apply  different completeness cuts  of 
$M_{\rm V}$~$\le$~$-$19.3 and $-$20.6.
More than 40\% of the bars that we detect have semi major axes 
$a <$~$0\farcs5$ and  would be easily missed in earlier 
surveys without the small point spread function of  ACS. 
The bars that we can reliably detect are fairly strong  
(with ellipticities $e \ge$~0.4) and have $a$ in the range  
$\sim$1.2--13 kpc.
We find that the optical fraction  of such strong bars 
remains at  $\sim$30\%~$\pm$~6\%  
from  the present-day out to look-back times  of 
2--6 Gyr ($z \sim$ 0.2--0.7) and 6--8 Gyr  ($z \sim$ 0.7--1.0); 
it certainly shows  no  sign of a  drastic decline at $z >$~0.7. 
Our findings of a large and similar bar fraction at these three 
epochs  favor scenarios in which cold gravitationally  
unstable disks  are already in place by $z\sim 1$ 
and where  on average bars have a long lifetime (well in excess of 2 Gyr).
The distributions of structural bar properties in
the two slices are,  however, not statistically identical and 
therefore allow  for  the possibility that the bar strengths and 
sizes may evolve over time.
\end{abstract}


\section{INTRODUCTION}

It is widely recognized that  stellar bars, either spontaneously 
or tidally induced,  redistribute mass and angular momentum
and thereby drive the dynamical and secular 
evolution of galaxies (e.g., Kormendy 1982; Shlosman, 
Frank, \& Begelman  1989; Pfenniger \& Norman 1990; 
Friedli \& Benz 1993; Kormendy \&  Kennicutt 2004). 
In the local universe, one-third  of 
local spiral galaxies host optically visible  strong bars 
(e.g., Eskridge \etal 2002, hereafter E02; see also $\S$ 3).  
Mounting evidence, including observations of 
central molecular gas   concentrations (e.g., Sakamoto \etal 1999), 
velocity fields (e.g., Regan, Vogel, \& Teuben 1997), 
and starbursts (e.g., Hunt \& Malkan 1999; 
Jogee, Scoville, \& Kenney 2004a), suggests that bars 
strongly influence their host galaxies.
However, the  most fundamental issues  
have yet to be resolved. 
When and how did bars form? Are bars a recent phenomenon or were 
they abundant at early cosmic epochs? 
Are bars  long-lived  or do they recurrently dissolve and re-form 
over a Hubble time? 
How do  stellar bars fit within the hierarchical 
clustering framework of galaxy evolution and  
relate to the underlying disk evolution?

The evolution of a bar over a Hubble time  depends on 
the host galaxy structure, the dark matter (DM) halo, and  
the environment.  Numerical simulations of this complex process 
make widely different predictions   (e.g.,  Friedli \& Benz 1993; 
Shlosman \& Noguchi 1993; El-Zant \& Shlosman 2002; Athanassoula 2002;  
Bournaud \& Combes 2002; Shen \& Sellwood 2004), while 
the handful of observational results  on bars  at intermediate 
redshifts  are conflicting. 
On the basis of 46 moderately inclined galaxies  imaged with the 
Wide Field Planetary Camera2 (WFPC2) in 
the Hubble Deep Field (HDF), Abraham \etal (1999, hereafter A99) claim 
a dramatic decline in the rest-frame optical bar fraction ($f_{\rm 
opt}$) from $\sim$24\% at $z\sim$~0.2--0.7 to below 5\%  at $z>$~0.7. 
On the basis of  Near-Infrared Camera and Multi-Object Spectrometer 
(NICMOS) images of 95 galaxies in the HDF  at $z\sim$~0.7--1.1,
Sheth \etal (2003, hereafter S03)  detect  four large bars  
with mean semi major axes $a$ of 12~kpc ($1\farcs4$), while 
smaller bars presumably escaped detection because of the large NICMOS 
point spread function (PSF).
S03 point out  that their observed  bar fraction of $\sim$5\% 
for large (12 kpc)  bars at $z >$ 0.7  is at least comparable to 
the local fraction of similarly large bars. 
From a study based on $Hubble Space Telescope$ ($HST$) 
Advanced Camera for Surveys (ACS) 
F814W images of 186 galaxies over a  $ 3\farcm9 \times  4\farcm2 $ 
area in the Tadpole field,  Elmegreen, Elmegreen, \& Hirst (2004,  
hereafter E04)  report a constant optical bar fraction of $\sim$20\%--40\%
at $z \sim$~0--1.1. This study is limited by the large 
(0.1--0.4) errors of the photometric redshifts (Ben{\'{\i}}tez \etal 2004), 
and the  small sample size precludes absolute magnitude  completeness  
cuts. Furthermore,  with only  a single ACS filter, the rest-frame 
bandpass of the observations shifts by more than a factor of 2 over 
$z \sim$~0--1.1.

In this Letter (see also Jogee \etal 2004b), 
we present  the first results of an extensive 
study of bars at  $z$~$\sim$~0.2--1.0, based on two-band 
ACS images covering  $14\arcmin \times 14\arcmin$  
($\sim$25\%) of  the Galaxy Evolution from Morphology and SEDs (GEMS) 
survey. The area  and the sample size of 1590 galaxies provide at least  
an order of magnitude better number statistics than earlier studies.  
We quantitatively identify bars using ellipse fits and 
minimize the effects of bandpass shifts by using both
F850LP and F606W images ($\S$ 2.2).
Using  highly accurate redshifts ($\S$ 2.1),  
we compare the bar fractions in two redshift slices after applying 
completeness criteria. We show that the  optical fraction of strong 
(ellipticities $e \ge$~0.4)  bars is remarkably constant ($\sim$ 30\%)  
from  the present-day out to look-back times  ($T_{\rm back}$) 
of 2--6 Gyr ($z \sim$~0.2--0.7)\footnote{We assume in this 
paper a flat cosmology with $\Omega_M = 1 - \Omega_{\Lambda} = 0.3$  
and $H_{\rm 0}$ =70~km~s$^{-1}$~Mpc$^{-1}$.} 
and 6--8 Gyr ($z \sim$~0.7--1.0).

\section{Observations, Sample, and Methodology}

\subsection{Observations and Sample Selection}
GEMS is a two-band (F606W and F850LP)  $HST$  ACS 
imaging survey (Rix \etal 2004)  
of  an 800 arcmin$^2$  ($\sim 28\arcmin \times 28\arcmin$) field
centered on the Chandra Deep Field-South.
GEMS consists of  78  one-orbit-long ACS pointings in each filter 
and reaches a  limiting 5 $\sigma$ depth for point sources 
of  28.3  and 27.1 AB mag in F606W and F850LP, respectively.
GEMS provides high-resolution ($\sim$$0\farcs5$
or  360 pc at  $z \sim$~0.7) ACS images for  $\sim$8300  
galaxies  in the redshift range  $z \sim$~0.2--1.1, 
where  accurate redshifts [$\delta_{\rm z}$/(1 + $z$)~$\sim$~0.02 
down to  $R_{\rm Vega}$ = 24] and spectral energy distributions  
(SEDs)  based  on 17 filters 
exist from the COMBO-17 project (Wolf \etal 2004). 
For this Letter, we analyze only $\sim$25\% of the GEMS  field 
as this area ($14\arcmin \times 14\arcmin$)  is already 
30 times that of the HDF and yields good number statistics and 
robust results on the bar fraction  ($\S$ 3). 
It provides a sample that consists of 1590 galaxies 
in the  range $z \sim$~0.2--1.0  and  $R_{\rm Vega} \le$ 24. 
In a future paper (S. Jogee \etal in preparation, hereafter Paper II), we 
will use the entire GEMS sample  to compare how bar 
properties evolve over 8 Gyr  at 1 Gyr intervals.

\subsection{Characterizing bars and $f_{\rm opt}$ out to $z\sim$~1.0}

Table 1 illustrates the two methods that we use for assessing 
bar properties  in two redshift slices at 0.25~$<z \le$~0.70 and  
0.7~$<z \le$~1.0. The first method  (referred to as  Method I in Table 1)  
is to identify bars in  the reddest filter (F850LP) at 
all $z$  in order to minimize extinction and better trace old stars. 
However, with this method,  the rest-frame bandpass shifts 
significantly, from $I$  to $B$ across the redshift range 0.2--1.0.
The second complementary method is to  trace bars in 
both F606W and  F850LP images  such that the rest-frame band 
remains relatively constant, between $B$ and $V$, out to $z \sim$~1.0.

We identify  bars and other galactic components in  F606W and  F850LP  
images via the widely used (e.g., Wozniak \etal 1995; 
Jogee \etal 1999, 2002; Knapen \etal 2000) 
procedure of fitting ellipses  using the standard IRAF  ``ellipse'' routine.  
We developed  a  wrapper that  automatically 
runs ``ellipse'' for a range of different  initial parameters, 
performing up to 200 fits per galaxy. 
We successfully fitted ellipses to 90\% of the 1590 galaxies, 
while the 10\% failure cases  included mostly very 
disturbed systems where no centering could be performed and  
some extended low surface brightness  systems. 
For all fitted galaxies, we inspected the image (Fig. 1a), the fitted 
ellipses overlaid on the images (Fig. 1b), and the radial  profiles (Fig. 1c) 
of intensity, ellipticity ($e$), and position angle  (P.A.) 
in order to confirm that  the best fit is  reliable. This 
inspection was aided by a  visualization tool that we developed. 
We identify a bar as such if the fitted ellipses and radial 
profiles show  the following characteristic bar signature (e.g, 
Wozniak \etal 1995)  illustrated 
in Fig. 1. (i)~The  ellipticity ($e$) must rise  to a 
\it global \rm  maximum $e_{\rm max}$, which we require to be
above 0.25, as well as above that of the outer disk, while the  P.A.   
has a plateau (within $\pm$~20$^\circ$) along the bar. 
(ii)~Beyond the bar end, as the bar-to-disk transition occurs, 
$e$ must drop by $\ge $~0.1, while the P.A. usually changes by 
$\ge$ $10^\circ$. 
From the profiles, we identify  $e$, P.A., and  semi major axes 
$a$ of both the  bar and the outer disk. 
We quantify the bar strength using $e_{\rm max}$, which  
correlates locally  with other measures of bar 
strength,  such as the gravitational bar torque 
(Laurikainen, Salo, \& Rautiainen 2002). 

The sample of 1430 galaxies with successful ellipse fits 
includes galaxies of different morphological types
(disks and spheroids), inclinations, and absolute magnitudes.
We apply two cuts at $M_{\rm V}$~$\le$~$-$19.3 and $-$20.6.  
The first cut gives us a sizeable sample of 
galaxies with a range of absolute magnitudes ($-$19.3 to $-$23.8)
similar to that of the 
Ohio State University (OSU) survey,  which is 
used to define the  local bar fraction  (EO2). 
However, $K$-corrections based on 
local Scd templates  (Coleman, Wu, \& Weedman 1980) suggest 
that we are only complete out to  $z \sim$~0.8 for the first
cut. The second more stringent cut at $-$20.6 ensures completeness out 
to $z \sim$~1.0, but it reduces the sample drastically  (see Table 1). 
In addition, to ensure reliable  bar detection, we use the disk 
inclination $i$ from ellipse fits ($\S$ 2.2) to 
exclude  highly inclined ($i > $~$60^\circ$) galaxies.

The optical bar fraction $f_{\rm opt}$ 
is  defined as ($N_{\rm bar}$/$N_{\rm sp/disk}$)  
where  $N_{\rm sp/disk}$ is the number of spiral or disk galaxies, 
and  $N_{\rm bar}$ is the number of such systems hosting bars.
We identify spiral/disk galaxies 
using three independent techniques (Table 1).
We first use the criterion  $n <$~2.5,  where 
$n$ is the index of  single-component 
Sersic models fitted to GEMS galaxies (B. H\"{a}ussler \etal 2004, in 
preparation)  with the GALFIT (Peng \etal 2002) package. 
Our choice of  $n <$~2.5 is dictated by the fact that a study
of GEMS galaxies at  $z\sim$~0.7 (Bell \etal 2004), 
as well as tests in which we  insert artificial galaxies in 
the GEMS fields, suggest that a Sersic 
cut of  $n \le 2.5$ picks up disk-dominated systems.
The second technique uses a cut $C <$~3.4, where $C$ is 
the  CAS (Conselice \etal 2000) concentration index.
Third, we use  rest-frame $U-V$  color cuts in the range 
0.8--1.2  to broadly separate spiral galaxies from red E/S0s, 
based on local SED templates  and 
the observed red sequence at  $z\sim$~0.7  (Bell \etal 2004).

\section{Results and discussion}

The bars that we identify primarily have ellipticities $e \ge$~0.4 
and  semi major axes $a$ in the range  $0\farcs15$--$2\farcs 2$ 
and 1.2--13 kpc (Fig. 2). 
Our experiment of artificially redshifting $B$-band images of a 
subset of OSU galaxies out to $z \sim$~1 shows that it is 
difficult to unambiguously identify  weaker ($e \le$ 0.3) bars, 
and we limit the discussion in this Letter to strong ($e \ge$~0.4) bars. 
Table 1  shows the optical bar fraction $f_{\rm opt}$  of such bars  
in the two redshift slices (0.25 $<z \le$0.70 and  0.7$<z \le$1.0)  
derived in the reddest filter and in  a relatively  fixed rest-frame
band ($\S$ 2.2). Results for  $M_{\rm V}$~$\le$~$-$19.3 are shown, 
based on 627 galaxies out of which we identify 258  moderately 
inclined ($i < $~$60^\circ$) spirals that host $\sim$80 bars.
The consistency in the six entries attests to 
the robustness of the results and shows that  \it \rm 
the  fraction of optically visible bars remains 
in the range  23\%--36\%  or at $\sim$30\%~$\pm$~6\% in both slices.
Incompleteness effects ($\S$ 2.2) do not bias the results since
the cut at $-$20.6  gives similar bar fractions, shown in 
brackets  (Table 1).
The bar fraction is slightly higher in the rest-frame $I$ band than 
$B$ band, possibly indicative of dust and star formation masking bars at 
bluer wavelengths. Our findings of a fairly constant  $f_{\rm opt}$
are consistent with E04 and do not show the dramatic decline in 
$f_{\rm opt}$ reported by A99.

More than 40\% of the bars that  we detect have semi major axes 
$a <$~$0\farcs5$  (Fig. 2), and many of these smallest bars may
have been missed in earlier WFPC2 (e.g., A99) and NICMOS studies
that did not benefit from the small ($0\farcs05$)  ACS  pixels,  
and the resulting narrow PSF. In addition to the wider WFPC2 PSF,
cosmic variance, low number statistics, and  methodology may have 
led to the lower $f_{\rm opt}$ reported by A99, but we cannot address 
this issue further here as the coordinates of galaxies in that 
study have not been published.

How do our results compare to $f_{\rm opt}$ for 
correspondingly  strong bars in the local universe? 
There are as yet no  statistics published on  $z \sim $~0 bars 
based on a  large volume-limited sample such as the Sloan Digitized 
Sky Survey. 
As the next best alternative, we turn to the OSU sample (E02). 
For $i <$~$60^\circ$ spirals, we find  
$f_{\rm opt}$~$\sim$~37\% for strong bars classified according 
to the  visual RC3 `SB' bar class, and $f_{\rm opt}$~$\sim$~33\% 
for strong bars classified according  to $e \ge$~0.4, where $e$ is 
based on ellipse fits to   OSU $B$-band images.
Thus, 
\it 
it appears that the optical fraction of strong ($e \ge$~0.4)
bars  remains remarkably similar  at $\sim$30\%--37\% 
from  the present-day out to look-back times  of 
2--6 Gyr ($z \sim$~0.2--0.7) and 6--8 Gyr  ($z \sim$ 0.7--1.0).
\rm
Our findings have several implications for disk and bar evolution.
(1)~The abundance of strong bars at early times implies that 
dynamically cold disks that can form large-scale stellar bars are already 
in place by $z\sim$~1.  They also suggest that 
highly triaxial, centrally concentrated DM halos, 
which tend to destabilize the bar (El-Zant  \& Shlosman 2002), 
may  not be prevalent in galaxies at $z \sim$~0--1.
(3)~The  remarkably similar fraction of  
strong bars at look-back times  of  6--8 Gyr, 2--6 Gyr, and in the present
day supports the view that large-scale stellar bars are long-lived (e.g.,
Athanassoula 2002; Shen \& Sellwood 2004; Martinez-Valpuesta \etal 
2004), with a lifetime  well above  2 Gyr.  
The alternative option, statistically allowed by the data, is that the
combined destruction and reformation timescale of bars is on average 
well below 2 Gyr. However, this is highly implausible because
the destruction of a bar leaves behind a dynamically hot disk that cannot
reform a bar unless it is substantially cooled via  processes that 
take at least several Gyr, such as the accretion of large amounts of 
cold gas. 
(3)~We note that a similar \it fraction \rm 
of bars at different  epochs  does not exclude 
the possibility that the bar strength (ellipticity) and 
its size can evolve in time because of intrinsic factors  and 
concurrent changes in the surrounding disk, bulge and halo 
potentials (e.g., Athanassoula 2002; Martinez-Valpuesta \& Shlosman 2004).
Kolmogorov-Smirnov  tests on the distributions of  $e$ and $a$ for 
different  magnitude cuts 
yield primarily  $P$ in the range  0.2--0.5, where $P$ is the significance 
level for the null hypothesis  that the two  data sets are drawn from 
the same distribution. Such values of $P$ hint at evolutionary 
effects, but  a larger sample is needed to draw definite conclusions. 
While this Letter focuses on the optical bar \it fraction, \rm 
in  Paper II, we will invoke the full GEMS sample 
of 8300 galaxies to  compare  bar 
and  host galaxy properties 
(e.g., disk scale lengths, masses, and $B/D$ ratios) over the last 
8 Gyr at 1 Gyr intervals.
 
\acknowledgments 
S.J., F.D.B., I.S., and  D.H.M. acknowledge support from  
NASA/LTSA/ATP grants  NAG5-13063, NAG5-13102, and ATP 5-10823, 
and NSF AST02-06251. E.F.B.  and S.F.S.  acknowledge  support from 
ECHPP under HPRN-CT-2002-00316, SISCO, and HPRN-CT-2002-00305, Euro3D RTN.
C.W.  was supported by a PPARC Advanced Fellowship. Support for  GEMS was
provided by NASA through GO-9500 from 
STScI, which is operated by 
AURA, Inc., for NASA, under NAS5-26555.

{}


\clearpage
\begin{deluxetable}{cccccccc}
\tabletypesize{\scriptsize}
\tablewidth{0pt}
\tablecaption{
Optical fraction $f_{\rm opt}$ of strong ($e \ge$0.4) bars at \newline
0.25~$<z \le$~0.70 ($T_{\rm back}$~$\sim$~2--6 Gyr) and 
0.7~$<z \le$~1.0 ($T_{\rm back}$~$\sim$~6--8 Gyr)
} 
\tablecolumns{7}
\tablehead{
\colhead {Redshift} &
\colhead{$N_{\rm gal}$} &
\colhead {Technique to} &
\colhead{$N_{\rm sp/disk}$} &
\colhead{Filter} &
\colhead{Rest-frame} &
\colhead{Optical bar} \\
\colhead{range} &
\colhead{} &
\colhead{identify } &
\colhead{} &
\colhead{to trace} &
\colhead{band} &
\colhead{fraction} \\
\colhead{} &
\colhead{} &
\colhead{disks/spirals} &
\colhead{} &
\colhead{bars} &
\colhead{} &
\colhead{$f_{\rm opt}$} & \\
\colhead {(1)} &
\colhead {(2)} &
\colhead {(3)} &
\colhead {(4)} &
\colhead {(5)} &
\colhead {(6)} &
\colhead {(7)} 
}
\startdata
\multicolumn{7}{c}{Method I: Using the  reddest ACS filter (F850LP) 
to identify bars}\\
0.25--0.70  & 384 (146) & Sersic $n$     & 148 (39)  & 
F850LP     & $I$ to $V$  &  36\% (33\%) \\	 
0.70--1.0    & 243 (135) & Sersic $n$    & 110  (49) & 
F850LP     & $V$ to $B$  &  24\% (27\%) \\	 
0.25--0.70  &    384 (146) & CAS $C$     & 170  (51) & 
F850LP     & $I$ to $V$  &  27\% (27\%) \\
0.70--1.0    &     243 (135) & CAS $C$   & 147  (73) & 
F850LP     & $V$ to $B$  &  23\% (23\%) \\
 0.25--0.70  &    384 (146) & $U-V$      & 175  (51) &
F850LP     & $I$ to $V$  & 29\%(28\%) \\
 0.70--1.0    &  243 (135) & $U-V$       & 139  (69) & 
F850LP     & $V$ to $B$  & 29\% (26\%) \\
\hline  
\multicolumn{7}{c}{Method II: Using F606W and F85OLP to identify 
bars at  an approximately  fixed rest-frame $B$/$V$ band} \\
0.25--0.70  & 384 (146)   & Sersic $n$ & 148 (39)  & 
F606W   &  $V$ to $B$ &   26\% (30\%) \\
0.70--1.0   & 243  (135)    & Sersic $n$ & 110  (49) & 
F850LP     & $V$ to $B$  & 24\% (27\%) \\	 
0.25--0.70  &  384 (146) &  CAS $C$      & 170  (51) & 
F606W   &  $V$ to $B$ &   24\% (28\%) \\
0.70--1.0    &     243 (135) & CAS $C$   & 147  (73) & 
F850LP     & $V$ to $B$  &  23\% (23\%) \\
0.25--0.70  &  384 (146) & $U-V$         & 175  (51) & 
F606W   &  $V$ to $B$ &  29\% (23\%) \\
 0.70--1.0    &   243 (135) & $U-V$      & 139  (69) &
F850LP     & $V$ to $B$  & 29\% (26\%) \\
\enddata 
\tablecomments{ 
Col. (1):  The redshift range. 
Col. (2): The no of galaxies in this range with 
$M_{\rm V}$~$\le$~$-$19.3 and $-$20.6 (shown in brackets) 
to which ellipses were fitted. 
Col. (3): The technique used to identify disk/spiral galaxies from 
the sample in (2). We use cuts of Sersic index $n <$~2.5,   
CAS concentration index $C <$~3.4, and rest-frame $U-V$~$<$~0.8--1.2.
Col. (4): $N_{\rm sp/disk}$, the  no of moderately inclined  
($i <$~$60^\circ$)  spiral/disk galaxies for the two magnitude cuts in (2).
Col. (5),(6): The filter and rest-frame band in which bars are traced; 
Col. (7):  $f_{\rm opt}$, the optical fraction  of strong bars, 
is  the fraction ($N_{\rm bar}$/$N_{\rm sp/disk}$) of 
moderately inclined spirals hosting bars with $e \ge$ 0.4.
Values for the two magnitude cuts of  $-$19.3 and $-$20.6 are shown.
}
\end{deluxetable}


\clearpage

\setcounter{figure}{0}
\begin{center}
\includegraphics[width=4.9in]{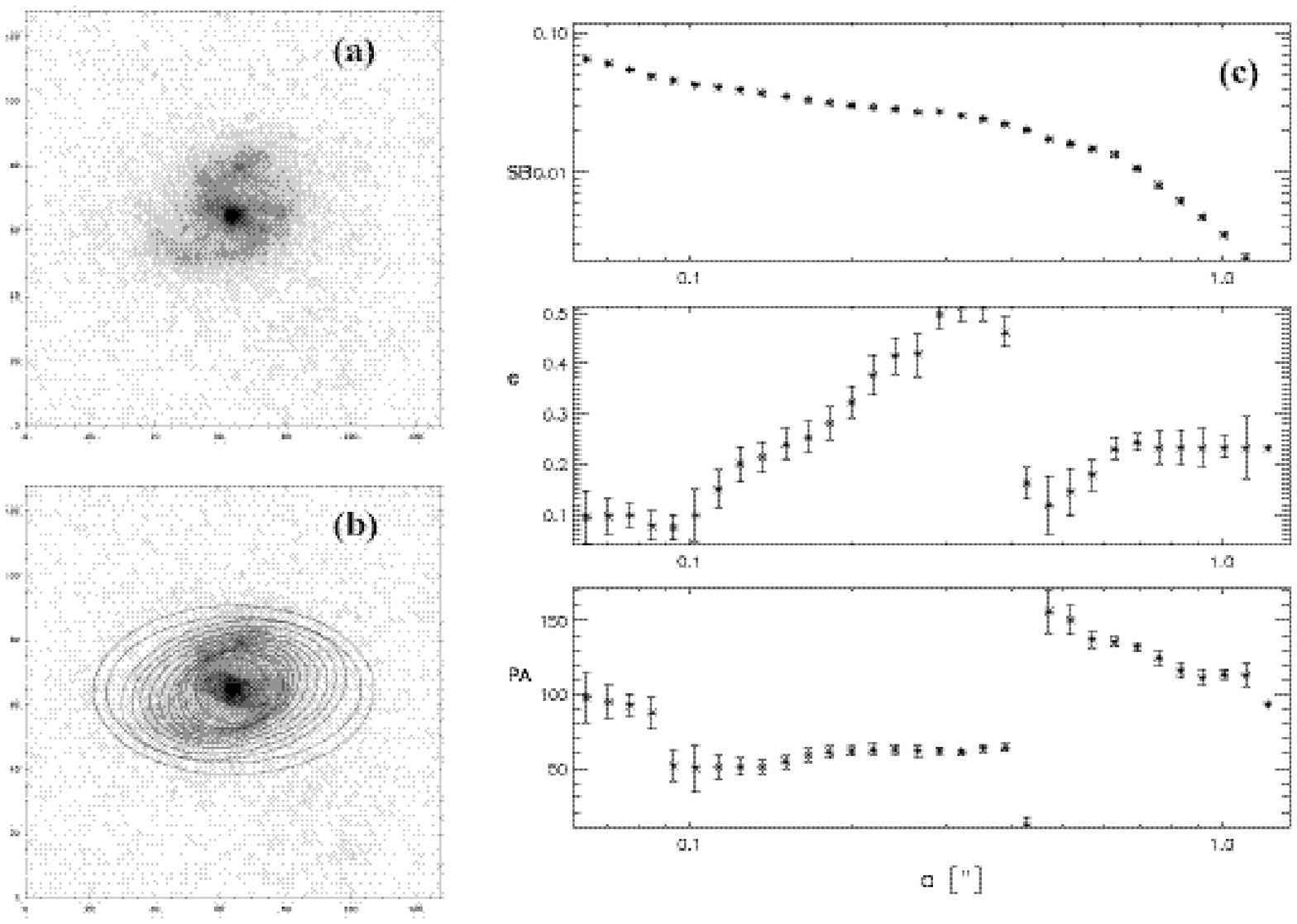}
\end{center}

\figcaption[fig1.ps]{  
Characterization of bars  out to $z \sim$~1.0: 
The  GEMS image of a  $z \sim$~0.5  galaxy with a  bar, prominent 
spiral arms, and a disk is shown without \it (a) \rm and with 
\it (b) \rm an overlay of the fitted ellipses.
\it (c) \rm In the resulting radial plots of the surface brightness,
ellipticity $e$, and P.A., the  bar causes $e$ to rise smoothly to 
a global maximum, while the P.A. has a plateau.
Beyond the bar end ($a$~$\sim$~$0\farcs36$), the 
spiral arms  lead to a twist in P.A. and varying $e$ before the
the disk dominates.}

%
\hspace{-7 mm}
\begin{center}
\includegraphics[height=2.5in]{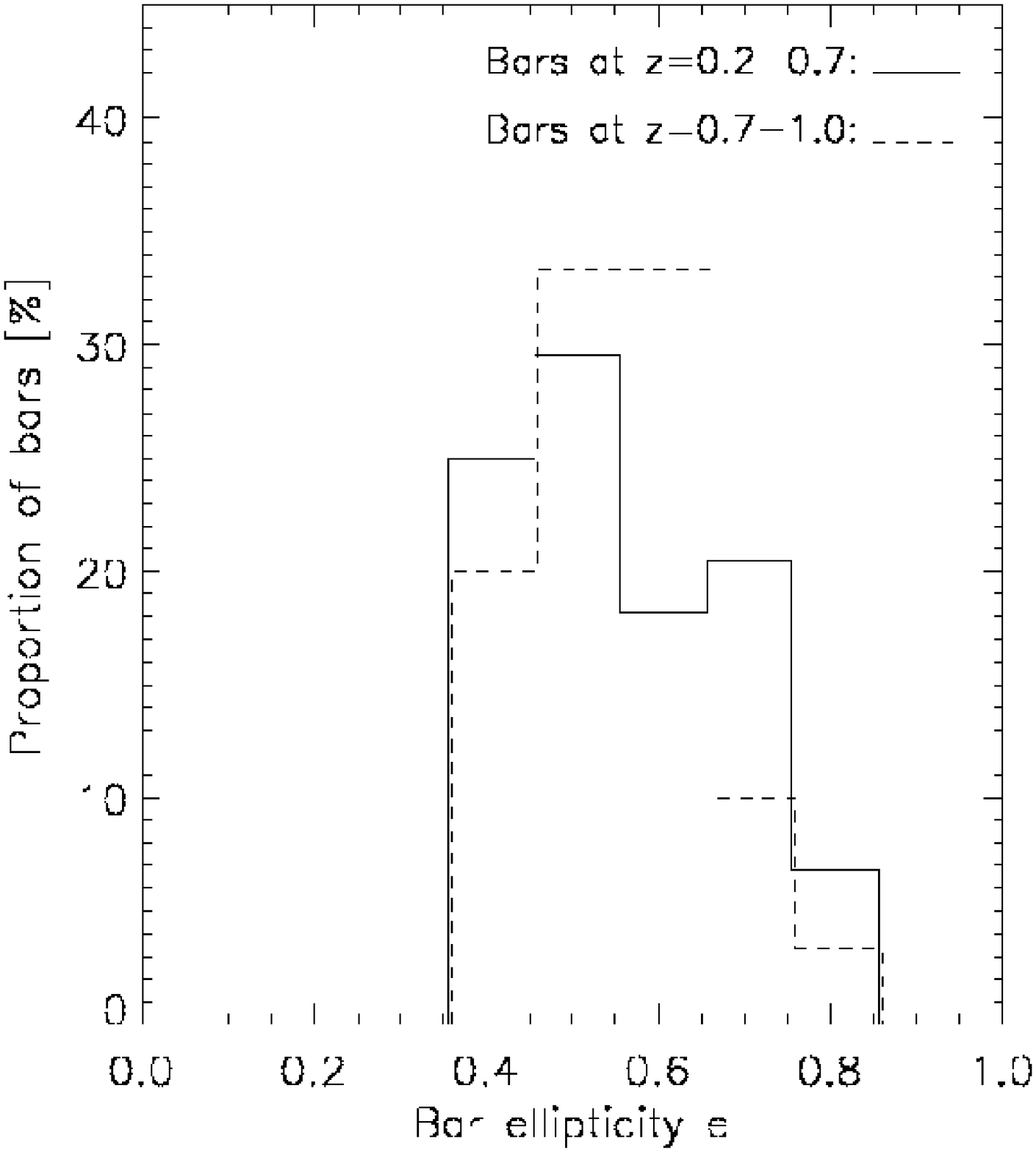}
\includegraphics[height=2.5in]{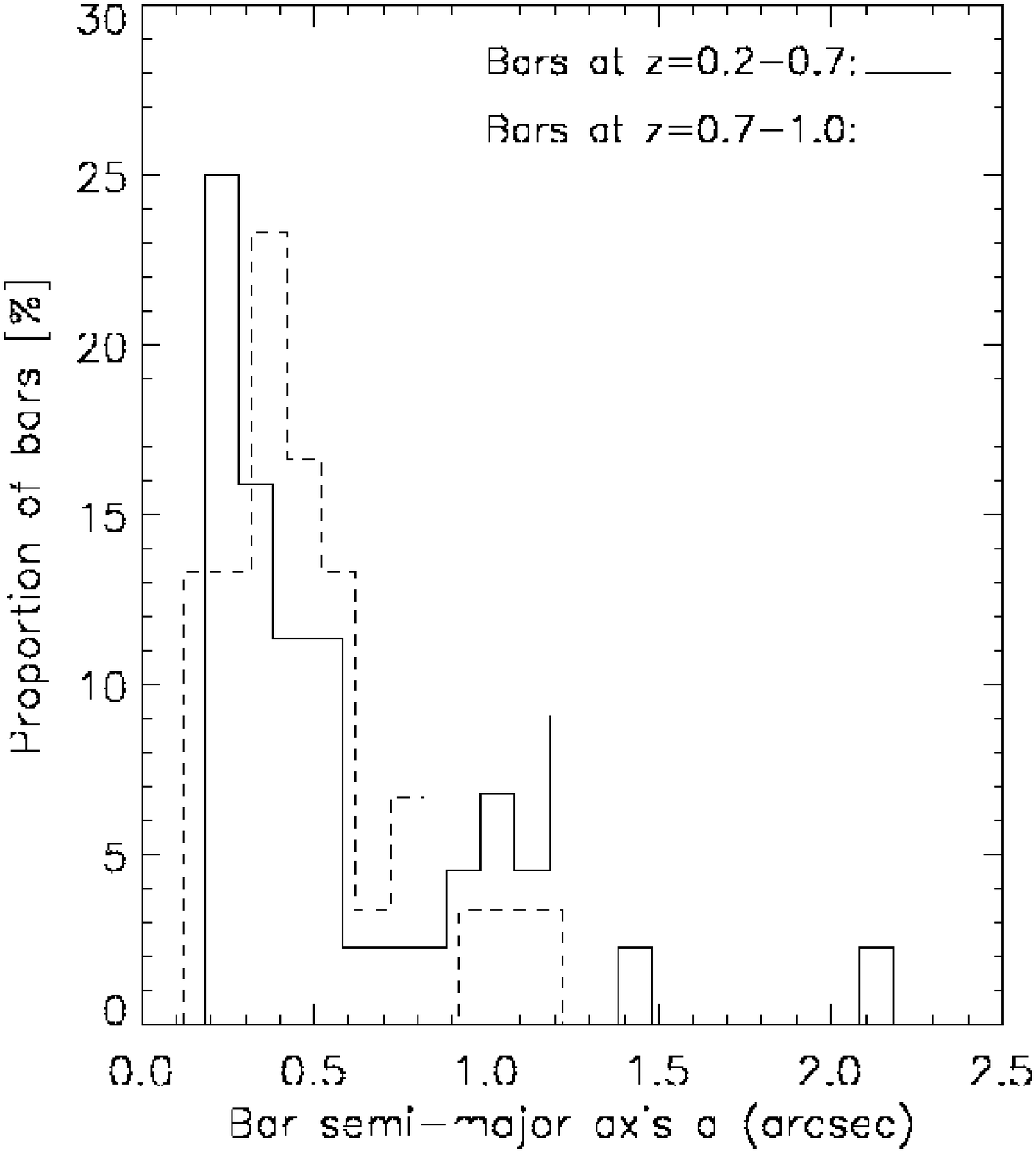}
\includegraphics[height=2.5in]{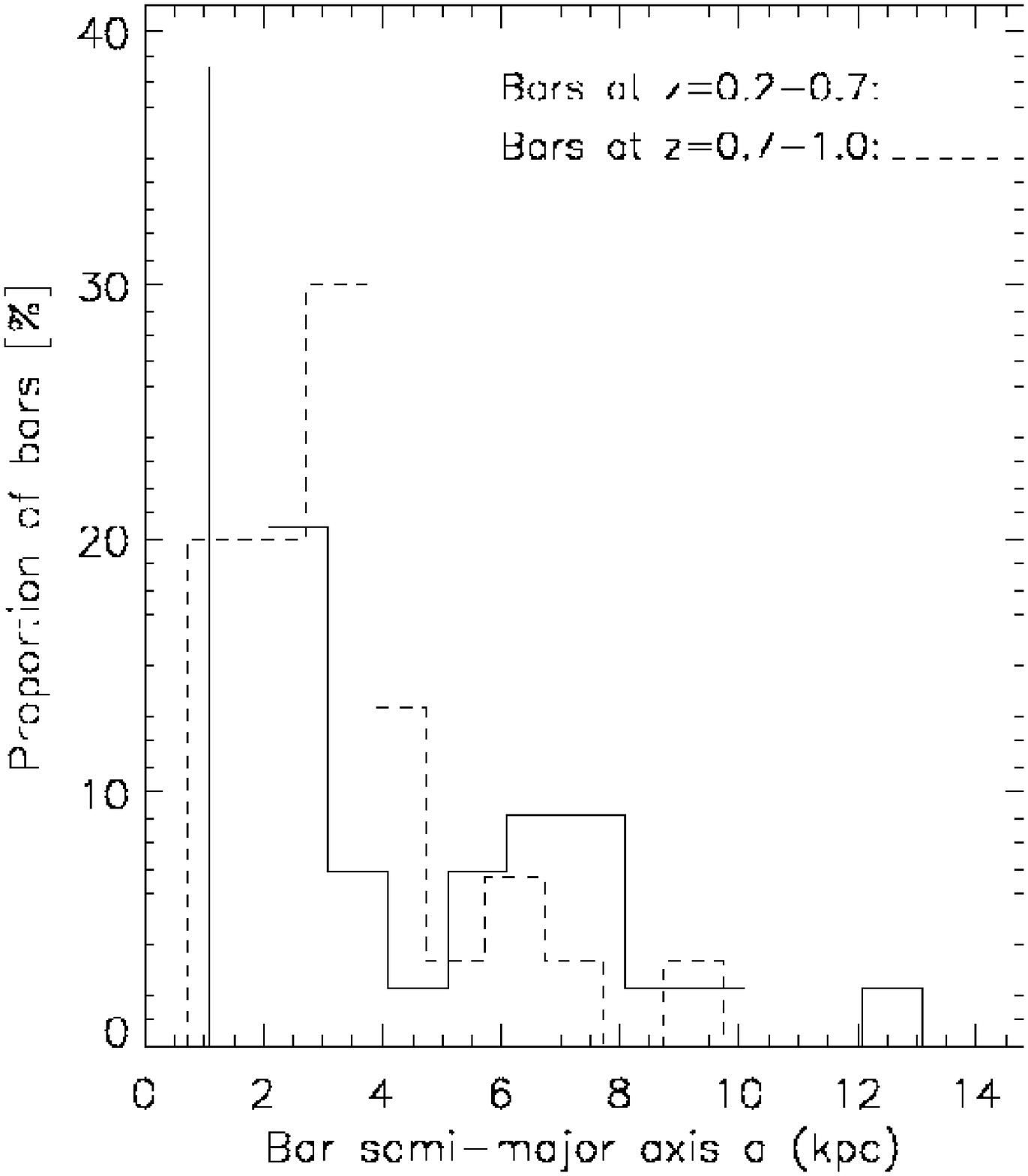}
\end{center}
\figcaption[fig2a.eps.fig2b.eps.fig2c.eps]{  
Comparison of  bars out to look-back times of  8 Gyr:  
The  bar  ellipticity $e$ \it (left) \rm  and semi major axis 
$a$ in $\arcsec$  \it (middle) \rm  and kpc \it (right) \rm are 
shown for bright   ($M_{\rm V}$~$\le$~$-$19.3), 
moderately inclined ($i <$~$60^\circ$) galaxies  at  
$z \sim$~0.2--0.7 ($T_{\rm back}$~$\sim$~2--6 Gyr) 
and $z \sim$~0.7--1.0 ($T_{\rm back}$$\sim$~6--8 Gyr).
The bars identified are primarily strong, with 
$e \ge $~0.4. A large fraction have $a <$~$0\farcs5$ and 
their detection is aided by the narrow ACS PSF.}

\end{document}